\documentclass[preprint2]{aastex}

\shorttitle{Sizes of galaxy halos in A1689}
\shortauthors{Halkola et al.}

\begin{document}

\title{The sizes of galaxy halos in galaxy cluster Abell 1689}

\author{A. Halkola and S. Seitz}
\affil{Universit\"ats-Sternwarte M\"unchen, Scheinerstra\ss e 1, D-81679 M\"unchen, Germany}
\email{halkola@usm.uni-muenchen.de}

\and

\author{M. Pannella}
\affil{Max-Planck-Institut f\"ur extraterrestrische Physik, Giessenbachstra\ss e, Postfach 1312, D-85741 Garching, Germany}

\begin{abstract}

  The multiple images observed in galaxy cluster Abell 1689 provide
  strong constraints not only on the mass distribution of the cluster
  but also on the ensemble properties of the cluster galaxies. Using
  parametric strong lensing models for the cluster, and by assuming
  well motivated scaling laws between the truncation radius $s$ and
  the velocity dispersion $\sigma$ of a cluster galaxy we are able to
  derive sizes of the dark matter halos of cluster galaxies.

  For the scaling law expected for galaxies in the cluster environment
  ($s\propto\sigma$), we obtain \mbox{$s = 64^{+15}_{-14} \times~(
  \sigma /~220~\mathrm{km/s} )$~kpc}. For the scaling law used for
  galaxies in the field with $s\propto\sigma^2$ we find \mbox{$s =
  66^{+18}_{-16} \times~( \sigma /~220~\mathrm{km/s}
  )^2$~kpc}. Compared to halos of field galaxies, the cluster galaxy
  halos in Abell 1689 are strongly truncated.

\end{abstract}


\section{Introduction}
\label{sec:intro}

Although galaxies are the units of objects seen on cosmological
distances, very little is known observationally about the extent of
dark matter halos surrounding the galaxies beyond the light emitted by
the gas and stars in them. Based on numerical simulations these dark
matter halos are expected to extend out to several hundred kpc
\citep[e.g. ][]{tormen:98}. Rotation curves of spiral galaxies and the
line of sight velocity dispersions of the stars in elliptical galaxies
can be measured only out to some tens of kiloparsecs (see
e.g. \citealt{sofue:01,bender:94} and references therein). The results
from these two methods indicate that the masses of galaxies continue
to grow roughly linearly with the radius, i.e. the matter in galaxies
is closely isothermal. The radial velocities of satellites of galaxies
can also be used to estimate the masses of their host galaxies
\citep[e.g. ][]{hartwick:78,zaritsky:89,prada:03}. This has recently
been done for the Milky Way by \citet{battaglia:05} who were able to
measure the radial velocity dispersion profile of the Galaxy out to
120~kpc. The method works only for field galaxies since it is
sensitive to other nearby massive galaxies and hence the galaxies
studied need to be isolated \citep{brainerd:04}.\\

Gravitational lensing is an ideal tool to measure the extents of dark
matter halos around galaxies since no optical tracers within the halo
are needed. Instead, the halo mass can be inferred from the
gravitational lensing of background sources.

Weak lensing can be used to study galactic dark matter halos
statistically. The field started from the pioneering work of
\citet{tyson:84}, and galaxy-galaxy lensing has now been successfully
used both in the field \citep{brainerd:96,dell'antonio:96,hudson:98,
fischer:00,smith:01,wilson:01,mckay:01,hoekstra:03,hoekstra:04} and in
clusters
\citep{natarajan:98,geiger:99,natarajan:02,gavazzi:04,limousin:06} to
measure the masses and extents of galaxy halos. The signal is very
weak for individual galaxies and needs to be collected from many
galaxies, possibly adopting various scaling laws to compare
measurements from lensing galaxies with different luminosities. The
different works generally find a tangential shear $\gamma$ that
decreases like $\gamma \propto 1 / \theta$ with the radius $\theta$,
i.e. the halos stay roughly isothermal beyond the luminous
component. In the field the signal from galaxies has been measured out
to $\sim$200~kpc \citep[e.g. ][]{wilson:01,hoekstra:04}.

The galaxy truncation in clusters has been studied both theoretically
and observationally in
\citet{natarajan:97,natarajan:98,geiger:98,geiger:99,natarajan:02,
gavazzi:04,limousin:05,limousin:06}. Strong truncation of galaxies is
found in \citet{natarajan:98}, \citet{natarajan:02} and
\citet{limousin:06} when compared to galaxies in the field (truncation
radii $s^*$ of $L^*$ galaxies span 17-55~kpc in the 6 clusters vs.
$s^*$=264$\pm 42$~kpc/$h_{70}$ in the field,
\citealt{hoekstra:04}). There is a general agreement that the halos of
galaxies in dense environments are truncated relative to those in the
field although the uncertainties are still large and the sample of
clusters used is inhomogeneous. The inherently statistical nature of
the methods and the need to assume certain scaling laws further
complicate the case. The method used in
\citet{natarajan:98,natarajan:02} also requires that the parameters of
the smooth cluster component are known accurately
\citep{natarajan:97}. This is achieved by incorporating also strong
lensing constraints in the clusters enabling \citeauthor{natarajan:02}
to accurately model the cluster profile.

The typical radius of an Einstein ring of a galaxy is at most a few
arcseconds and so in the strong lensing regime it is not possible to
probe the extent of dark matter halos beyond a few arcseconds
directly. This makes strong lensing unfeasible to study the dark halos
of individual galaxies beyond a few arcseconds in the field. In
clusters of galaxies however the combined potentials of the smooth
dark matter halo of the cluster as a whole and those of the individual
galaxy halos are responsible for the lensing of background
sources. This enables us to statistically probe galaxy halos in dense
environments using strong gravitational lensing. Since lensing
constrains only the total potential of the cluster it is important to
have a large number of multiple images with a large radial spread over
the cluster in order to investigate the different mass components
separately. Abell 1689 is ideally suited for this task with the large
number of identified multiple images of many different background
sources and well defined strong lensing models
\citep{broadhurst:05,diego:05b,zekser:06,halkola:06}. In this paper we
use the strong lensing models developed in \citet{halkola:06} to study
the truncation of cluster galaxy dark matter halos in A1689. We
demonstrate that the models are indeed sensitive to the total mass
contained in the cluster galaxies and derive sizes for the galaxies in
the cluster. This is the first time the sizes of galaxy halos have
been measured in dense cluster environments with strong lensing
only.\\

In section \ref{sec:SLmodel} we give a brief summary of the models
used in \citet{halkola:06} in particular the modeling of the galaxy
component of the cluster, in section \ref{sec:methodology} we outline
the methodology used to study the truncation of the cluster
galaxies. The results are presented in section \ref{sec:results} and
in section \ref{sec:checks} we perform several checks to demonstrate
that the results obtained are robust and reasonable. In section
\ref{sec:comparison} we compare the results with earlier published
studies of galaxy halo truncation before concluding in section
\ref{sec:conclusion}.\\

The cosmology used throughout this paper is $\Omega_{m}$=0.30,
$\Omega_{\Lambda}$=0.70 and H$_{0}$=70~km/s/Mpc.

%__________________________________________________________________
%__________________________________________________________________
%__________________________________________________________________
%__________________________________________________________________

\section{Strong Gravitational Lensing Model for A1689}
\label{sec:SLmodel}

The strong lensing models in this work are based on the parametric
models used in \citet{halkola:06} to study the mass profile of A1689
in detail. Here we give a short summary of the strong lensing modeling
but refer the reader to \citet{halkola:06} for the details.

The multiple images in \citet{halkola:06} were in most part those
identified in \citet{broadhurst:05}. In total a 107 multiple images in
31 multiple image systems and 1 arc were used. In 5 cases at least one
of the images in a system had also a spectroscopic redshift and the
redshifts were kept fixed for these systems. The redshifts of another
26 multiple image systems were estimated using photometric
redshifts. In these cases the redshift of an image system was allowed
to find its best redshift within the estimated photometric redshift
errors. The arc was too faint for good photometry and its reshift is
left unconstrained.

The mass in the cluster is assumed to be in two smooth DM halos that
are described by either non-singular isothermal ellipsoids (NSIE) or
elliptical Navarro-Frenk-White profiles (ENFW). The small scale mass
structures associated with the galaxies are modeled with BBS profiles
\citep{brainerd:96}. The BBS profile is a singular isothermal sphere
with a truncation radius $s$. In the central regions ($r<s$) the
density profile is isothermal ($\rho \propto r^{-2}$) but there is a
truncation of the halo at radius $s$ after which the density falls
sharply with $\rho \propto r^{-4}$.

The velocity dispersions of the galaxies are estimated using the
Fundamental Plane (FP). The FP ties kinematic (velocity dispersion),
photometric (effective surface brightness) and morphological (half
light radius) galaxy properties together
\citep{dressler:87,djorgovski:87,bender:92}. Measuring morphological
and photometric properties of the galaxies allows us to estimate the
galaxy kinematics. We assume that the central velocity dispersions of
galaxies, as derived from the FP, are equal to the halo velocity
dispersions, and that the masses in disks can be neglected. For some
fainter galaxies we have also used the Faber-Jackson relation
\citep[here after FJ relation]{faber:76} that relates the absolute
magnitude of a galaxy to its velocity dispersion. The FJ relation has
a large intrinsic scatter and the velocity dispersion obtained using
FJ have a larger uncertainty than the ones obtained with the FP.

The truncation radii of the galaxies are assumed to follow a scaling
law of the form $s_{gal} = s^0 \times~( \sigma_{gal} / \sigma^0 ) ^
{\alpha}$. In this paper we discuss the same scaling laws as in
\citet{halkola:06}, namely $\alpha=1$ and $\alpha=2$. $\alpha$=1
corresponds to tidal truncation of halos in dense cluster environment
\citep{merritt:83} whereas galaxies with $\alpha$=2 have a constant
mass-to-light ratio and is usually assumed in weak lensing analyses
\citep[e.g. ][]{brainerd:96,natarajan:98,hoekstra:04}. In this paper
we explore further the radial extent of the galaxy halos for the
scaling laws used in \citet{halkola:06}.

A1689 is an excellent candidate for this work since the large number
of multiple images ensures not only that the global mass profile can
be constrained very accurately but also the relative contributions of
the smooth DM and galaxies can be determined as will be shown later.

\section{Methodology}
\label{sec:methodology}

In this paper take advantage of the unique opportunity presented in
A1689 to use strong lensing and the significant contribution of the
cluster galaxies on the positions of the impressive number of multiple
images observed in the cluster.The effect is only observable in the
total fit quality and is hence statistical in nature in that extension
of individual galaxies cannot be determined.

Unlike the usual galaxy-galaxy lensing in which foreground galaxies
weakly distort the shapes of background galaxies this method relies on
the changes induced by galaxies on the \emph{positions} of multiply
imaged background galaxies. This strong galaxy-galaxy lensing is only
applicable in the strong lensing regime where multiple images are
observed over a large range of cluster centric radii so that they pose
strong constraints both on the total cluster potential and also on the
galaxies.

In this section we outline the method used to measure the extents of
galaxy halos. The strong lensing models are constrained by the
observed multiple images. The positions of the images can be measured
to an accuracy of better than 1 pixel or 0.05'' on the images from the
Advanced Camera for Surveys. The only other measurables are the
redshifts of the cluster and the multiple images. The redshift of the
cluster is well established from spectroscopic surveys
\citep{teague:90,balogh:02,duc:02} and the overall mass scale of the
cluster is fixed by the 5 spectroscopic redshifts of multiple image
systems. The major uncertainty in the models is the inclusion of the
cluster galaxies. In the following we describe the Monte-Carlo
simulations used to find the normalization of the scaling law, $s^0$,
and how these simulations can also be used to estimate the error in
$s^0$ due to the uncertainties in the observables.

\subsection{Monte-Carlo Simulations}
\label{sec:methodology:simulations}

A Monte-Carlo run consists of reassigning a new velocity dispersion,
$\sigma_{MC}$, to each cluster galaxy based on the value
$\sigma_{gal}$ and estimated error determined using the FP or the FJ
relation. The new $\sigma_{MC}$ of a galaxy was drawn from a Gaussian
distribution centered on $\sigma_{gal}$ with a width corresponding to
the estimated error. The multiple image positions were similarly
varied with assumed error of 1 pixel. In this way we have constructed
a simulated galaxy that has properties similar to the one observed
within our estimates of the errors.

This cluster can now be analyzed in the same way as the 'original'
cluster. This means that we find the optimal parameters for the two
smooth DM halos (positions, ellipticities, position angles and the two
free parameters of the halos: velocity dispersion and core radius for
the NSIE profile and concentration and virial radius for the ENFW
profile) and redshifts for the image systems with photometric
redshifts.

The simulated clusters used in this work are the same that were used
in \citet{halkola:06} to derive errors for the total mass profile and
the parameters of smooth DM halo.

In this work we concentrate on the normalization of the truncation
radius $s^0$ for two scaling laws, $\alpha=1$ and $\alpha=2$, which
was not done in \citet{halkola:06}. This means that in addition to
optimizing the above mention parameters we also find the optimal value
of $s^0$ for each simulated cluster. This is explained below.

\subsection{Determining $s^0$ for a Monte-Carlo Run}
\label{sec:methodology:s0_one}

The optimal $s^0$ for each Monte-Carlo run was taken as the one with
the minimum $\langle~\chi^2 \rangle^{1/2}$ when $s^0$ was
progressively increased from 20~kpc to 200~kpc. The parameter
optimization was performed by a source plane minimization for
computational reasons. In all subsequent analysis we have used an
image plane $\chi^2$ defined as the sum of the squared distances
between the observed images and ones predicted by our
models. $\langle~\chi^2 \rangle^{1/2}$ is hence the rms distance
between the observed images and the corresponding model image
positions.

We use the shapes of these $\langle~\chi^2 \rangle^{1/2}$ vs. $s^0$
curves to convince the reader that there is clear signal and that
$s^0$ can be constrained in clusters using strong lensing once
sufficiently many multiple images can be used to constrain the models.

\subsection{Determining $s^0$ for the Cluster}
\label{sec:methodology:s0_cluster}

The shape and spread of the $\langle~\chi^2 \rangle^{1/2}$ vs. $s^0$
curves could in principle also be used to derive confidence limits on
$s^0$. This, however, would require us to perform more simulations to
derive appropriate $\Delta \chi^2$ levels for the confidence
limits. The $\langle~\chi^2 \rangle^{1/2}$ vs. $s^0$ curves do however
demonstrate that there is a strong and clean signal that can be used
to derive $s^0$ and the errors for a given scaling law.

The best fitting $s^0$ and the errors for the cluster are derived from
the distribution of the $s^0$ values obtained in the Monte-Carlo runs
instead.

\section{Results}
\label{sec:results}

For the scaling law we need to choose a reference $\sigma^0$. The
derived $s^0$ is then the truncation radius of a galaxy with a
velocity dispersion equal to this $\sigma$. The truncation radii of
galaxies with different $\sigma$s can then be obtained using the
appropriate scaling law. In this work we simply assume a fiducial
value of $\sigma^0$=220~km/s. To compare the $s^0$ obtained in this
work with literature one should scale our $s^0$ by \mbox{(
$\sigma^0_{lit}$ / 220~km/s ) $^{\alpha}$}.\\

In creating the simulated clusters the velocity dispersions for the
galaxies are drawn from a Gaussian distribution and hence we do not
expect to see significant differences in the shapes of the individual
$\langle~\chi^2 \rangle^{1/2}$ vs. $s^0$ curves between the different
Monte-Carlo runs that would arise from a systematic change in the
galaxy component. The curves do vary in their absolute $\langle~\chi^2
\rangle^{1/2}$ level however. For this reason we have normalized the
individual curves to their respective median $\langle~\chi^2
\rangle^{1/2}$ in order to bring all the curves to a similar
$\langle~\chi^2 \rangle^{1/2}$ level. After this the curves have been
renormalized to the level of the mean median $\langle~\chi^2
\rangle^{1/2}$ of all the curves. The scaling of the individual curves
is necessary in order to combine the information on $s^0$ from the
different curves.\\

\begin{figure*}
  \centering
  \includegraphics[height=2.8in]{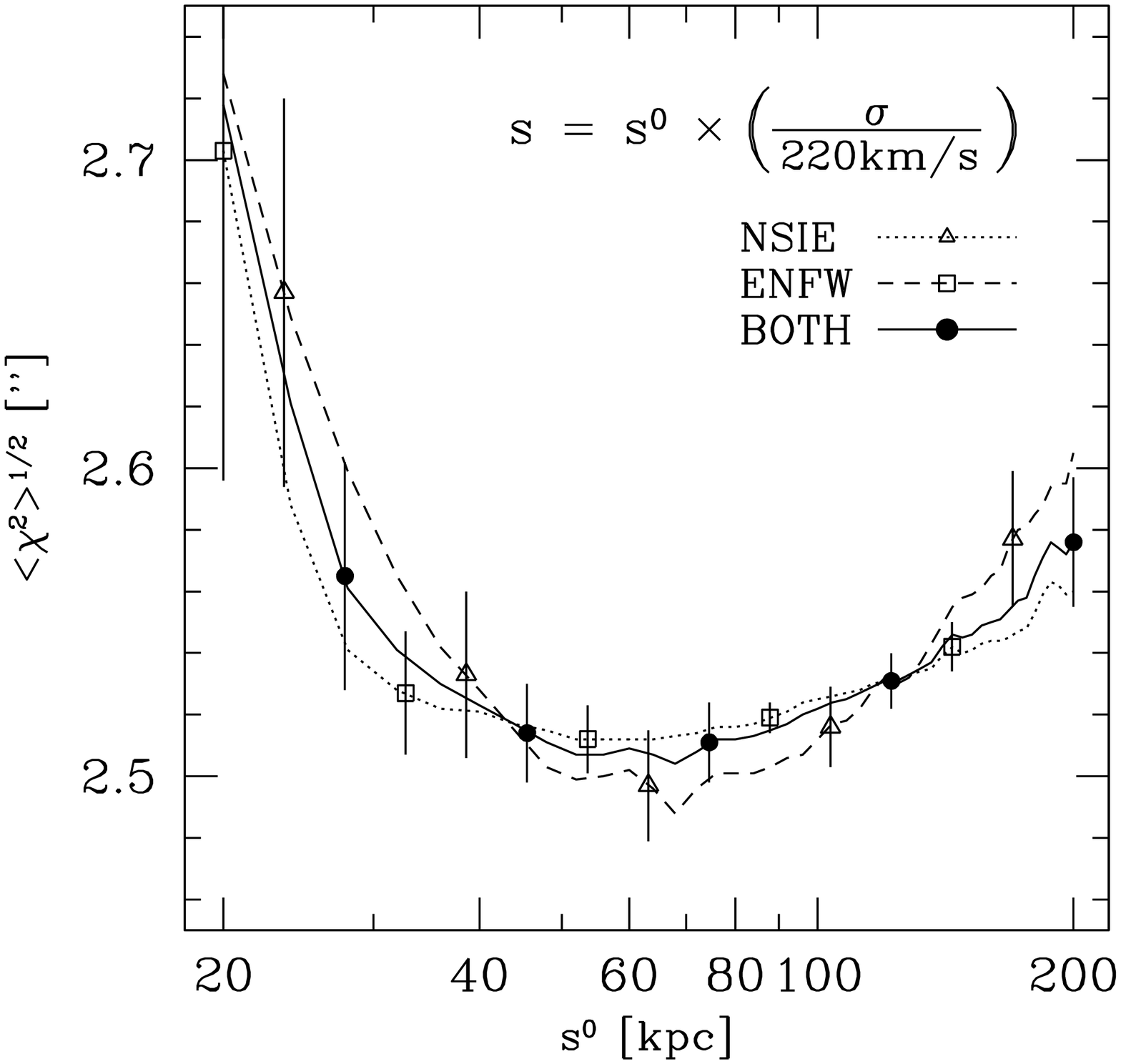}
  \includegraphics[height=2.8in]{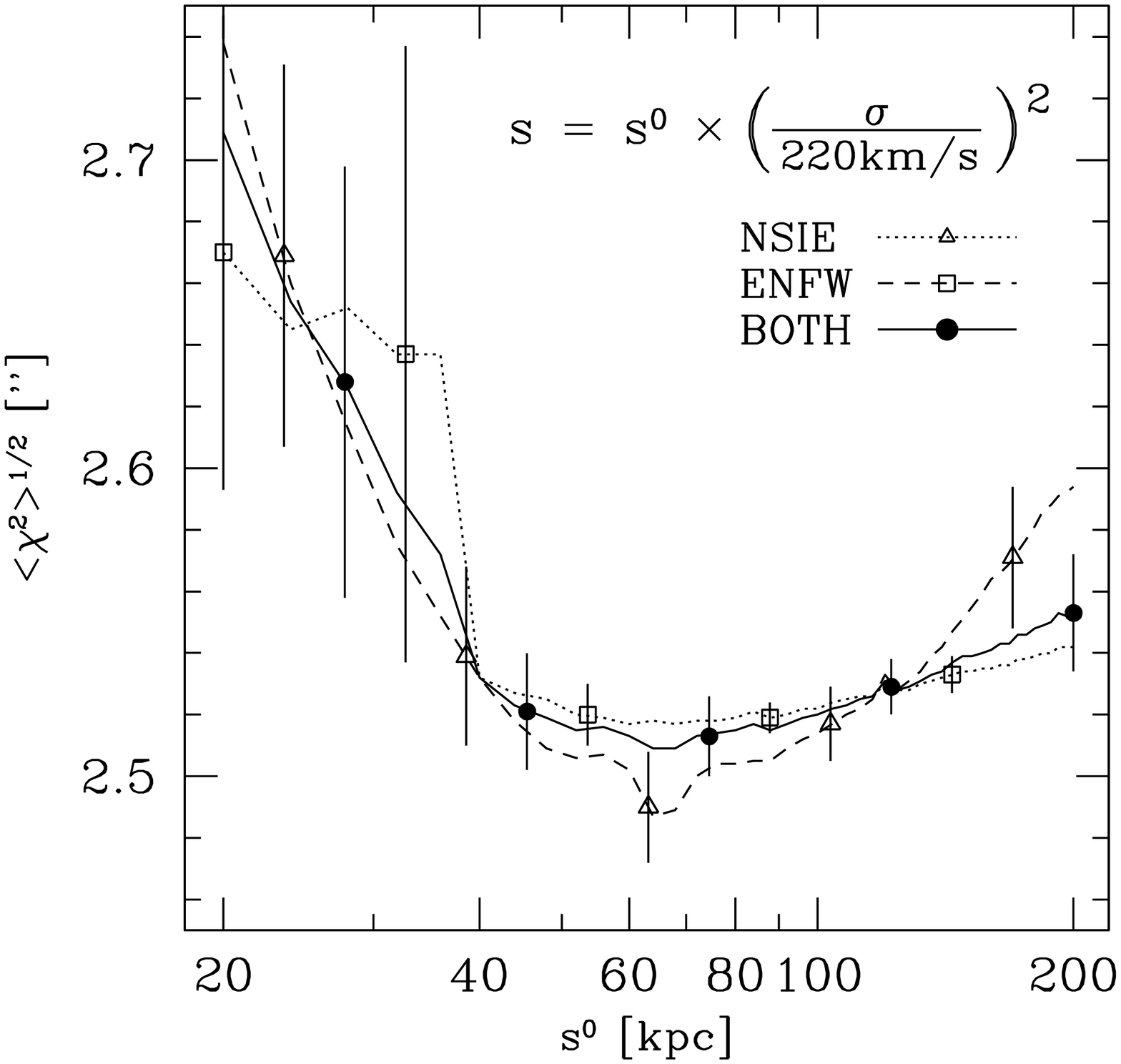}\\
  \caption{The mean $\langle~\chi^2 \rangle^{1/2}$ vs. $s^0$ curve for
    1000 simulated clusters for the smooth DM described by both NSIE
    (dotted line, squares) and ENFW (dashed line, triangles), and the
    combined data from the two smooth profiles (solid line,
    circles). The left panels are calculated for the scaling of the
    truncation radius with $s \propto \sigma$ and the right panels for
    $s \propto \sigma^2$. The points show the 2-sigma clipped mean for
    each $s^0$, and the error bars show the final sigma of the clipped
    points. Before clipping, the individual curves were normalized to
    their median $\langle~\chi^2 \rangle^{1/2}$ in order to bring all
    the curves to a similar $\langle~\chi^2 \rangle^{1/2}$ level for
    comparison. The median curve has been brought back to the level of
    the mean median $\langle~\chi^2 \rangle^{1/2}$. The combined curve
    has been calculated from the NSIE and ENFW curves and not from the
    individual curves for the two models. The minimum $\langle~\chi^2
    \rangle^{1/2}$ is obtained at $\sim$60-70~kpc.}
  \label{fig:results1}
\end{figure*}

In Fig. \ref{fig:results1} the mean curves for 1000 simulated clusters
for each of the smooth DM profiles used are shown. NSIE is shown as a
dotted line and squares, ENFW as a dashed line and triangles. The left
panel shows the curves for $\alpha=1$ ($s \propto \sigma$) and the
right for $\alpha=2$ ($s \propto \sigma^2$). Combining the two smooth
DM profiles yields the solid line and points shown as circles. The
points show the 2-sigma clipped mean for each $s^0$. The error bars
show the dispersions of the final clipped points for a given $s^0$.

The smaller scatter in the points for NSIE models is an indication
that the renormalized curves are very similar while the considerable
scatter for the ENFW models shows that the curves differ not only in
the absolute $\chi^2$ level but also in their shape. The combined
curve has been calculated from the NSIE and ENFW curves and not from
the curves of each individual Monte-Carlo run for the two models. ENFW
smooth DM profiles generally favor slightly larger values for $s^0$
than the NSIE models ($\Delta s^0\sim$~10~kpc).\\

\begin{figure*}
  \centering
  \includegraphics[height=2.8in]{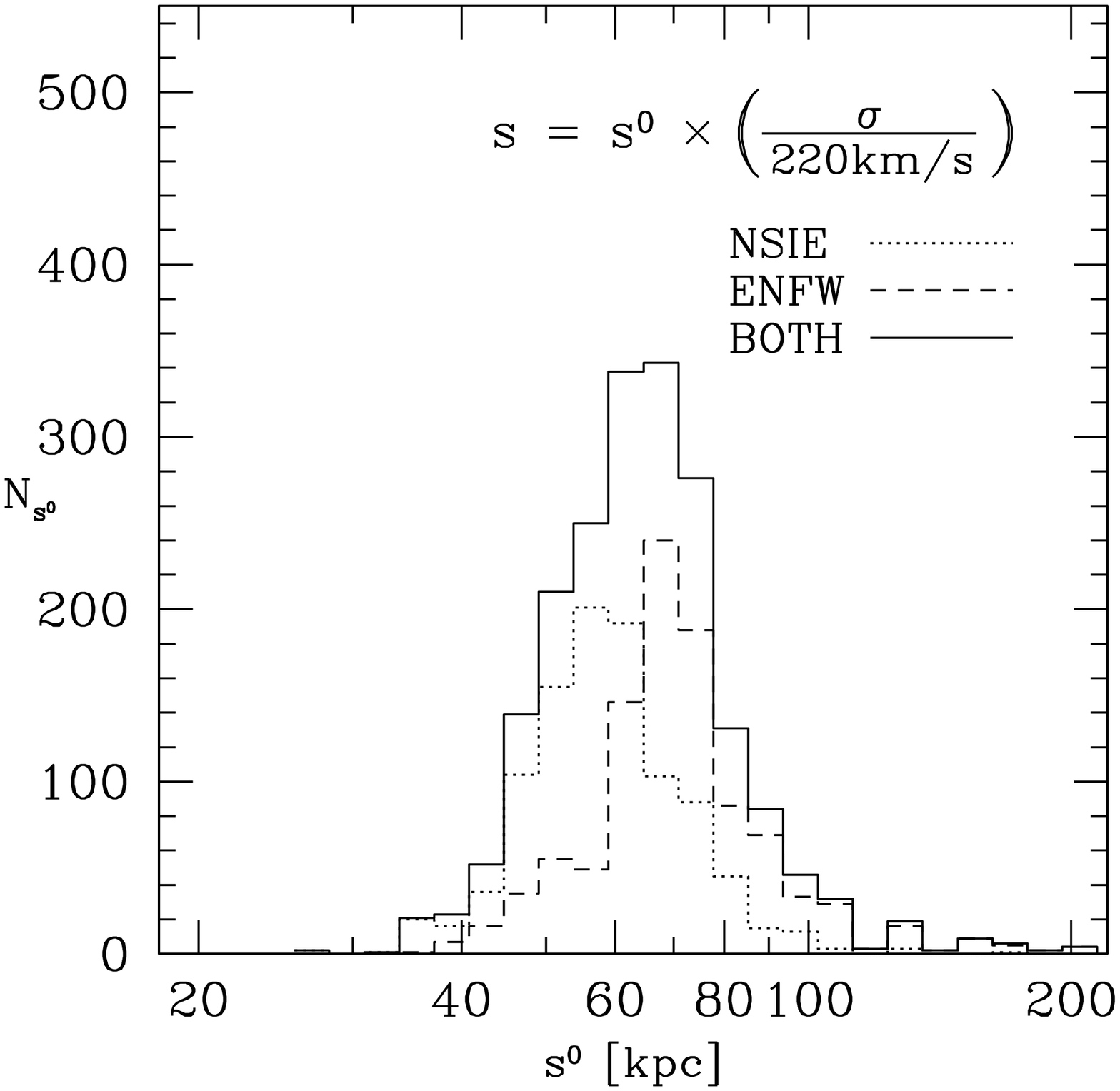}
  \includegraphics[height=2.8in]{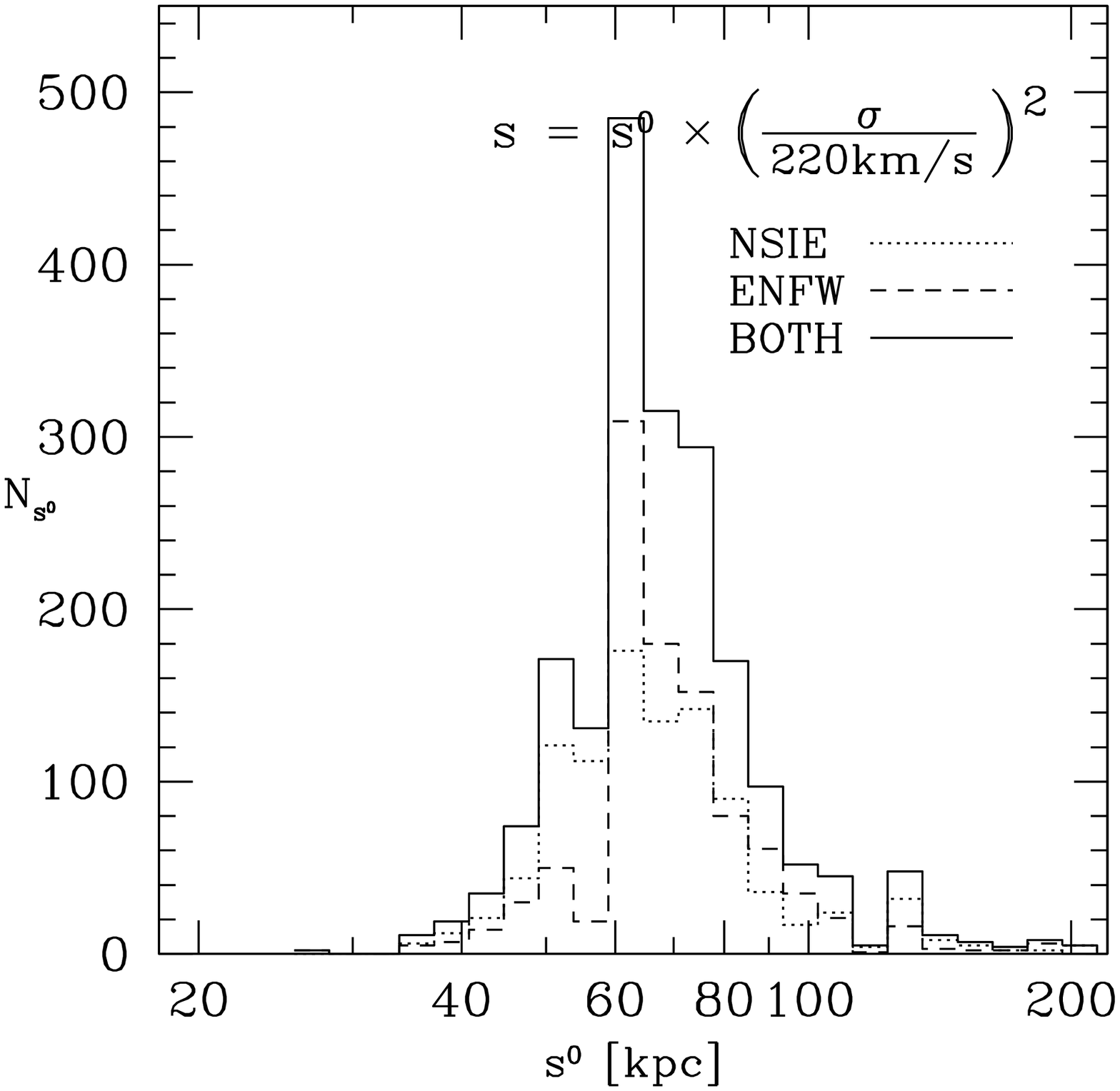}
  \caption{The histograms of the $s^0$ values at which each simulated
    cluster attains its minimum. The line types used are as above. As
    is expected, the histograms peak nicely at the positions where the
    mean curves on the top row also have their minima. The strong
    clustering of the histogram between $s^0$~=~40~kpc and
    $s^0$~=~90~kpc with only a few outliers demonstrates that $s^0$ is
    strongly constrained. The best fit $s^0$ and its error has been
    derived from these histograms.}
  \label{fig:results2}
\end{figure*}

The $\langle~\chi^2 \rangle^{1/2}$ vs. $s^0$ curves are flatter at
large ($>$50~kpc) $s^0$ than at smaller $s^0$, this is especially
apparent for the cases where the smooth DM is described by an NSIE
profile. A possible reason for the shallower slope on the logarithmic
horizontal scale (linear in fractional change in mass) in
Fig. \ref{fig:results1} is that the larger extent of the halos, and
hence a smoother combined mass profile of the galaxies, makes it
easier for the smooth DM component to compensate for the change in the
mass in the galaxies. In the small $s^0$ regime the galaxies have
significant local contribution to the image positions which cannot be
easily compensated by the smooth DM component.\\

The curves for the two scaling laws are very similar and we are not
able to differentiate between them in terms of quality of fit. This is
also seen in weak lensing determinations of the extensions of the dark
matter halos of galaxies (Limousin 2006, private communication). A
possible explanation is that instead of measuring the extension
directly we are in fact measuring the mass of the galaxies. This then
creates a degeneracy between the two parameters in the scaling law,
namely $\alpha$ and $s^0$. For a different value of $\alpha$ the same
total mass in the galaxies can be obtained by appropriately adjusting
$s^0$.\\

We would like to stress at this point that the two panels in
Fig. \ref{fig:results1} are only shown to illustrate that the $s^0$ is
indeed constrained and to provide an idea how the fit quality changes
when $s^0$ deviates from the best fit $s^0$. The $\langle~\chi^2
\rangle^{1/2}$ vs. $s^0$ curves are not used to derive the $s^0$ of
the cluster nor the errors. The estimation of $s^0$ and errors is done
using the histograms of the $S^0$s obtained using the Monte-Carlo
simulations shown in Fig. \ref{fig:results2} and explained below.\\

We finally use the best fitting $s^0$ of each Monte-Carlo run to
derive $s^0$ for A1689 and estimate the errors. In Fig.
\ref{fig:results2} we show the histograms of $s^0$ values at which
each simulated cluster attains its minimum $\langle~\chi^2
\rangle^{1/2}$, i.e. the best fitting $s^0$ for a given Monte-Carlo
run.

As is expected the histograms peak nicely at the positions where the
mean curves in Fig. \ref{fig:results1} also have their minima. The
strong clustering of the histograms between $s^0$~=~40~kpc and
$s^0$~=~90~kpc with only a few outliers demonstrates that $s^0$ is
well constrained. The flatter $\langle~\chi^2 \rangle^{1/2}$ vs. $s^0$
curves at $s^0>$50~kpc for the NSIE models lead to less well defined
minima and correspondingly wider distribution of $s^0$ at large $s^0$
in the histograms.

The best fit values of $s^0$ for the different descriptions of the
smooth DM component of the cluster are shown in Table
\ref{tab:data}. The values given are the geometric means of the best
fit $s^0$ of all the simulated clusters. We have used the geometric
mean to estimate the truncation radius since this corresponds to
fractional change in mass and the $\langle~\chi^2 \rangle^{1/2}$
vs. $s^0$ curves in Fig. \ref{fig:results1} are relatively symmetric
in $\log(s^0)$ (although not exactly as discussed earlier). We also
give in Table \ref{tab:data} the estimated 1- and 2-$\sigma$ errors of
$s^0$. The errors have been derived from the distribution of the best
fit $s^0$ of the simulated clusters shown in
Fig. \ref{fig:results2}. For this the histograms were interpreted as
probability distributions of $s^0$ and 1- and 2-$\sigma$ confidence
intervals were estimated by the regions around the mean that contain
68.3 per cent and 95.4 per cent of the best fit $s^0$ values from the
simulation for the 1- and 2-$\sigma$ errors respectively.  The
asymmetry of the distribution becomes evident at higher confidence
limits as can be seen in the 2-$\sigma$ errors.

For the scaling law expected theoretically for galaxies in clusters
\citep{merritt:83}, \mbox{$s = s^0 \times~( \sigma /~\sigma^0 )$}, we
find $s^0=64^{+67}_{-28}$~kpc, where the errors given are 2-$\sigma$
errors. For the scaling law used for galaxies in the field (\mbox{$s =
s^0 \times~( \sigma /~\sigma^0 )^2$}), we find
$s^0=66^{+72}_{-26}$~kpc, the errors are again 2-$\sigma$ errors.\\

\begin{table*}
  \centering
  \caption{ Derived $s^0$ values and 1- and 2-$\sigma$ errors for
    $s^0$ for the different descriptions of the smooth DM component of
    cluster. The truncation radius s of a galaxy depends on its
    velocity dispersion $\sigma$ and the scaling laws adopted are of
    the form $s = s^0 \times~( \sigma / \sigma^0 ) ^ {\alpha}$. The
    $s^0$ values given are the geometric means of the individual
    minima of the simulated clusters. The errors are derived from the
    distribution of the minima. We give both 1- and 2-$\sigma$ errors
    since the asymmetries of the distributions become more apparent at
    higher confidence limits. The histograms of the minima for the
    different descriptions of smooth DM are shown in
    Fig. \ref{fig:results2}.}
  \vspace{2mm}
  \begin{tabular}{ccccc}
    \hline
    \bf{Smooth DM}&&$\mathbf{s^0}$&\bf{1-$\sigma$ errors}&\bf{2-$\sigma$ errors}\\
    \bf{profile}&$\mathbf{\alpha}$&\bf{(kpc)}&\bf{(kpc)}&\bf{(kpc)}\\
    \hline
    \hline
    \bf{NSIE}        & 1 & 58 & +12 / --11 & +32 / --23 \\
    \bf{ENFW}        & 1 & 69 & +19 / --12 & +88 / --30 \\
    \bf{NSIE \& ENFW}& 1 & 64 & +15 / --14 & +67 / --28 \\
    \hline
    \bf{NSIE}        & 2 & 64 & +18 / --15 & +76 / --25 \\
    \bf{ENFW}        & 2 & 66 & +18 / --16 & +70 / --26 \\
    \bf{NSIE \& ENFW}& 2 & 66 & +18 / --16 & +72 / --26 \\
    \hline
  \end{tabular}
  \label{tab:data}
\end{table*}

\section{Checks on the Robustness of the Results}
\label{sec:checks}

We have performed the following checks to confirm that the results
presented above are reasonable and robust.

\subsection{$\sigma^0$, $s^0$, $\alpha$ and the Total Mass in Cluster Galaxies}
\label{sec:checks:mass}

The total mass of a galaxy with a BBS profile can be easily written in
terms of its truncation radius $s$ and velocity dispersion $\sigma$ as
is shown in \citet{brainerd:96}. The total mass of the cluster
galaxies in A1689 with a scaling law for the truncation of the halos
of the from $s = s^0 \times~( \sigma / \sigma^0 ) ^ {\alpha}$ can be
written simply as,

\begin{equation}
  \label{eqn:mtot}
  M_{tot}=7.3 \times 10^{5}~\bigg( {s^0 \over \mathrm{kpc}} \bigg)~
  \sum_{i} \bigg({\sigma_{i} \over \mathrm{km/s} } \bigg)^2 ~ \bigg(
  \frac{\sigma_{i}}{\sigma^0}\bigg)^{\alpha} ~\mathrm{M_{\sun}},
\end{equation}

where $s_{0}$ is the normalization of the scaling law, $\sigma^0$ a
reference velocity dispersion and $\sigma_i$ is the velocity
dispersion of galaxy i.

In our study we have taken $\sigma^0$~=~220~km/s. Note that this
$\sigma^0$ is only a fiducial value and is not related to the L$^*$ of
the galaxies in the cluster. With this $\sigma^0$ and our set of
galaxies in the cluster, the galaxies have the same total mass with
the two scaling laws ($\alpha$=1 and $\alpha$=2) if
\mbox{$s^0_{\alpha=2} = 0.93~\times~s^0_{\alpha=1}$}.  Note that this
relation between $s^0_{\alpha=1}$ and $s^0_{\alpha=2}$ is the same for
all $s^0_{\alpha=1}$ and $s^0_{\alpha=2}$.

The fact that the normalizations $s^0$ obtained for the two scaling
laws are very similar ($s^0_{\alpha=1}=64^{+15}_{-14}$~kpc and
$s^0_{\alpha=w}=66^{+18}_{-16}$~kpc) for $\sigma^0$=220~km/s provides
strong support for the results and our analysis.

\subsection{Sensitivity of Cluster Lensing to Extensions of Galaxy Halos}
\label{sec:checks:mock}

To demonstrate that we are indeed able to measure the extension of
galaxy DM halos with strong lensing we have created clusters with an
$s^0$ in the range [20,80]~kpc. For each of these clusters we have
created a mock set of multiple images that are exactly reproduced by
the cluster. The mock multiple image set is based on the observed
multiple images so that the cluster setup is as close to reality as
possible. These clusters with known galaxy truncation laws are then
analyzed in the same way as is done for A1689.

We find that we are able to recover the input $s^0$ within a few kpc
in all cases. Additionally, both the change in the fit quality
($\Delta \langle~\chi^2 \rangle^{1/2}\sim$0.2'') of these new
simulated clusters and the distribution of the best fit $s^0$ is
similar to what is observed and shown in Figures \ref{fig:results1}
and \ref{fig:results2}.

\subsection{Effect of the Choice of Multiple Image Systems}
\label{sec:checks:multipleimages}

We have in addition checked the sensitivity of the results to our
choice of multiple images. This was done by running another 100
Monte-Carlo runs with $\alpha$=1 and the smooth DM described by NSIE
profiles. This time for each Monte-Carlo run we selected randomly 20
of the 32 image systems to use as constraints for the modeling. The
multiple images with spectroscopic redshifts were always included
since they are needed to fix the overall mass scale of the
cluster. With fewer constraints we obtained essentially the same $s^0$
with larger spread in the distribution of $s^0$ from the different
runs. The best fit $s^0$ obtained with 20 image systems is $s^0 =
59^{+27}_{-19}$~kpc compared to $s^0 = 58^{+12}_{-11}$~kpc with all
the image systems, the errors are 1-$\sigma$.

When only 20 multiple image systems were used the absolute
$\langle~\chi^2 \rangle^{1/2}$ stayed at the same level as with all
the 32 image systems. This shows that the $\langle~\chi^2
\rangle^{1/2}$ level is not driven by only a few image systems but all
image systems contribute similarly to the $\langle~\chi^2
\rangle^{1/2}$ level. The change in fit quality between best fit $s^0$
and extrema at $s^0$=20~kpc and $s^0$=200~kpc with fewer image systems
is $\Delta \langle~\chi^2 \rangle^{1/2}\sim$0.1'' showing that also
the individual $s^0$ are less well constrained with fewer image
systems. That no change in $s^0$ is obtained demonstrates that our
results for $s^0$ are robust.

\section{Comparison with literature}
\label{sec:comparison}
%__________________________________________________________________
%__________________________________________________________________

The extensions of dark matter halos have been measured previously in
cluster environment by \citet{natarajan:98}, \citet{natarajan:02},
\citet{gavazzi:04}, and \citet{limousin:06}.

Strong truncation of galaxies is found in \citet{natarajan:98},
\citet{natarajan:02}, and \citet{limousin:06} when compared to
galaxies in the field; the truncation radii of an L$_*$ galaxy span
the range 17-55~kpc for the 6 clusters studied in
\citet{natarajan:02}. The halos in \citet{limousin:06} are truncated
more with a typical truncation radius below 20~kpc.

An important difference in the analysis of
\citet{natarajan:98,natarajan:02} to that of \citet{limousin:06} is
that \citeauthor{natarajan:02} also include strong lensing features in
the central parts of the clusters to further constrain the mass
profile of the selected cluster sample and hence also constrain the
galaxy halo parameters stronger. This helps to better define the shear
contribution from the cluster galaxies and hence the truncation
radius. Another major difference is that \citet{limousin:06} work
exclusively with ground based data where as Hubble Space Telescope
data are used in \citet{natarajan:98,natarajan:02}.

The large errors in the work of \citeauthor{gavazzi:04} are caused by
the smoothing scale of $\theta_s$=220~kpc/$h_{70}$ employed in their
analysis which restricts the achievable resolution. Although they are
not able to derive strong limits on the sizes of cluster galaxies,
they do find that halos on the periphery of the cluster MS0302+17 are
more strongly truncated than the halos on the central regions of the
cluster providing thus further confirmation for the tidal stripping
scenario.\\

\begin{figure}[t]
  \centering \plotone{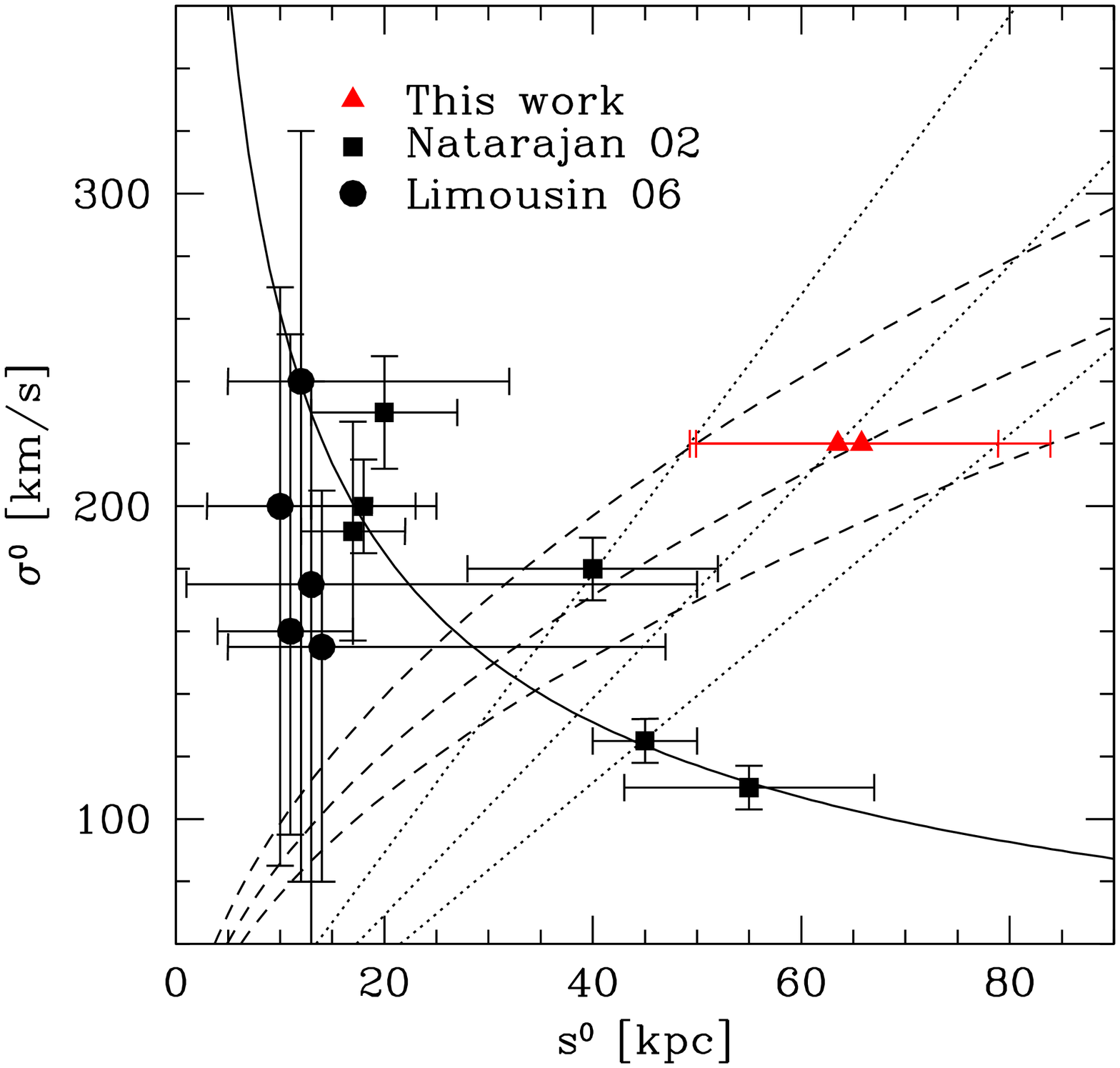}
  \caption{A comparison of three studies of galaxy truncation in the
  dense cluster environments. The red triangles are from this work,
  squares from \citet{natarajan:02} and the circles from
  \citet{limousin:06}. The error bars shown are all 1-$\sigma$. The
  dotted lines show the scaling of the best fit values and errors for
  $\alpha=1$, the dashed lines for $\alpha=2$. With the solid line we
  show the $s-\sigma$ pairs for a galaxy with a total mass of
  5$\times$10$^{11}\mathrm{M_{\sun}}$.}
  \label{fig:clus_comp}
\end{figure}

For the range of $\sigma^0$s for the clusters in \citet{natarajan:02}
our $s^0$ is in the range [32,66]~kpc for $\alpha=1$ and [16,72]~kpc
for $\alpha=2$ (their $s^0$ span 17-55~kpc). In
Fig. \ref{fig:clus_comp} we show a comparison between our results and
those of \citet{natarajan:02} and \citet{limousin:06}. For our points
we also show the scaling of $s^0$ with $\sigma^0$ as dotted and dashed
lines ($\alpha=1$ and $\alpha=2$ respectively). The lines can be used
to convert the $s^0$ and errors to a $\sigma^0$ different from
220~km/s, making the comparison between other works easier. The solid
line shows the $s-\sigma$ pairs for a galaxy with a total mass of
5$\times$10$^{11}\mathrm{M_{\sun}}$. The scatter of the points is
large, though mostly consistent within the large error bars. There is
some indication that the galaxy halos in A1689 are more extended than
those in most of the other cluster studied.

\citet{natarajan:02} compared their results for the density of the
cluster at the core radius, $\rho(r_c)$, and the truncation radius of
galaxies obtained in their analysis and found results in good
agreement with \citet{merritt:83},

\begin{equation}
  \label{eqn:tidal}
  s^0\ =\ 40 \ \frac{\sigma^0}{180 \mathrm{km/s}} \left(
  \frac{\rho(r_c)}{3.95 \times 10^{6} \mathrm{M_{\odot} kpc^{-3}}}
  \right)^{-0.5}\mathrm{kpc.}
\end{equation}

Using the results for A1689 \citep[$\sigma_{cluster}$=1450~km/s and
$r_c$=77~kpc, ][]{halkola:06} we get an expected truncation radius of
54~kpc, a little smaller than the $\sim$65~kpc obtained in this
work. This (small) difference is in fact also expected since in our
analysis we measure the truncation radii of the galaxies along the
line of sight. Some of the galaxies will have large cluster-centric
distances despite their small projected distances from the
center. This supports the idea that the galaxy clusters are mainly
truncated by the tidal field of the global potential as assumed by
\citet{merritt:83} and also shown in numerical simulations by
\citet{moore:98,ghigna:00}.\\

When comparing the results from different works it should be noted
that weak lensing works generally include all the galaxies from the
center to the periphery of the cluster (although \citealt{gavazzi:04}
do separate the galaxies in radius). This means that the results are
averaged over the cluster galaxy population out to several Mpc
\citep{limousin:06}. With our strong lensing method we include
galaxies only out to a projected cluster-centric radius of
$r\sim300$~kpc. The clusters also vary in their central densities
complicating direct comparison between clusters. According to
\citet{limousin:06} their cluster sample (Abell clusters A1763, A1835,
A2218, A383 and A2390) form a homogeneous set of clusters and hence
the results for these clusters should be comparable.

\begin{figure}[t]
  \centering
  \plotone{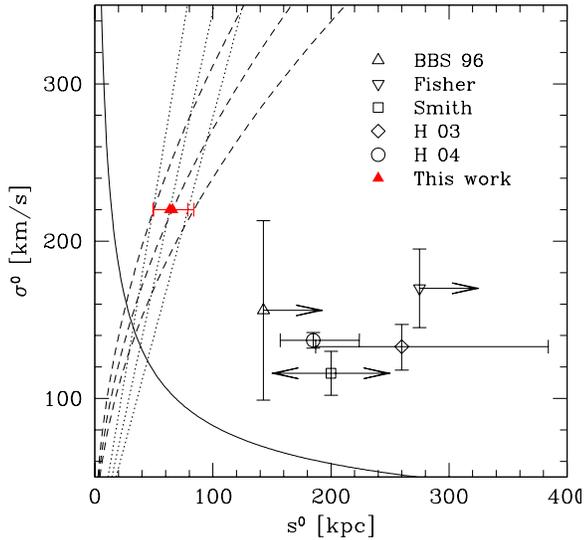}
  \caption{A comparison of studies of galaxy truncation in the
  field. The red triangles are from this work. The dotted lines show
  the scaling of the best fit values and errors for $\alpha=1$, the
  dashed lines for $\alpha=2$. The discrepancy is clear between the
  results obtained for galaxies in clusters and those in the
  field. Notice the difference in the horizontal scale between this
  figure and Fig. \ref{fig:clus_comp}. The solid line shows the
  $s-\sigma$ pairs for a galaxy with a total mass of
  5$\times$10$^{11}\mathrm{M_{\sun}}$.}
  \label{fig:field_comp}
\end{figure}

Comparison to field galaxies is shown in Fig. \ref{fig:field_comp}. In
the figure we show points from \citet{brainerd:96},
\citet{fischer:00}, \citet{smith:01}, \citet{hoekstra:03} and
\citet{hoekstra:04}. Adopting $\sigma^0_{136}$=136~km/s used by
\citet{hoekstra:04} we obtain $s^0_{136}=39^{+41}_{-17}$~kpc for
$\alpha$=1 and $s^0_{136}=25^{+25}_{-10}$~kpc for
$\alpha$=2. Similarly to previous studies of cluster galaxies we
report a strong truncation of galaxy halos in dense cluster
environments compared to galaxy halos in the field.

\section{Summary and Conclusions}
\label{sec:conclusion}
%__________________________________________________________________
%__________________________________________________________________

In this paper we report the determination of the sizes of galaxy dark
matter halos in galaxy cluster A1689. The strong lensing models for
the cluster are constrained by 107 multiple images and an arc in 32
image systems. The strong constraints from these images enable us to
study not only the global mass profile of the clusters but also the
ones of the cluster galaxies. Assuming well motivated scaling laws
between the truncation radius of a galaxy halo and its central
velocity dispersion (as obtained with the fundamental plane and
Faber-Jackson relations) we can study the combined effect of the
cluster galaxies on the multiple images and the ensemble properties of
the galaxies. This is the first time the sizes of galaxy halos have
been measured using strong lensing only.

For a scaling law of the form $s_{gal} = s^0 \times~( \sigma_{gal} /
\sigma^0 ) ^ {\alpha}$ we find $s^0=64^{+67}_{-28}$~kpc for $\alpha=1$
and $s^0=66^{+72}_{-26}$~kpc for $\alpha=2$. Both values are given for
a fiducial galaxy velocity dispersion of $\sigma^0$=220~km/s. The
errors are 2-$\sigma$ errors to show the clear asymmetry of the
errors. The $s^0$s are in good agreement with previously determined
values in several other clusters using weak lensing
\citep{natarajan:98,natarajan:02,limousin:06}.

Galaxy halos in a cluster can be truncated either by the tidal field
of the global cluster potential or harassment
\citep{moore:96,moore:98} by other cluster galaxies that strip the
halos of galaxies in the central regions of cluster. Mergers of
cluster galaxies on the other hand are extremely rare
\citep{ghigna:98}. Once the cluster has been formed the principal
mechanism for truncation is the tidal stripping of galaxy halos by the
global cluster potential \citep{ghigna:00}. This is supported by the
correlation between the density of the cluster at the core radius and
the truncation radii of galaxies shown in \citet{natarajan:02}. The
results presented here also support the tidal stripping scenario.

\acknowledgments

This work was supported by the Deutsche Forschungsgemeinschaft, grant
\emph{SFB 375} ``Astroteilchenphysik''. We would like to thank
P. Schneider for giving us useful comments on the manuscript and
M. Limousin for providing us with a draft of \citet{limousin:06} and
for interesting discussions.

{\it Facilities:} \facility{HST (ACS)}, \facility{HST (WFPC2)}.


\begin{thebibliography}{42}
\expandafter\ifx\csname natexlab\endcsname\relax\def\natexlab#1{#1}\fi

\bibitem[{{Balogh} {et~al.}(2002){Balogh}, {Couch}, {Smail}, {Bower}, \&
  {Glazebrook}}]{balogh:02}
{Balogh}, M.~L., {Couch}, W.~J., {Smail}, I., {Bower}, R.~G., \& {Glazebrook},
  K. 2002, MNRAS, 335, 10

\bibitem[{{Battaglia} {et~al.}(2005){Battaglia}, {Helmi}, {Morrison},
  {Harding}, {Olszewski}, {Mateo}, {Freeman}, {Norris}, \&
  {Shectman}}]{battaglia:05}
{Battaglia}, G., {Helmi}, A., {Morrison}, H., {Harding}, P., {Olszewski},
  E.~W., {Mateo}, M., {Freeman}, K.~C., {Norris}, J., \& {Shectman}, S.~A.
  2005, MNRAS, 364, 433

\bibitem[{{Bender} {et~al.}(1992){Bender}, {Burstein}, \& {Faber}}]{bender:92}
{Bender}, R., {Burstein}, D., \& {Faber}, S.~M. 1992, ApJ, 399, 462

\bibitem[{{Bender} {et~al.}(1994){Bender}, {Saglia}, \& {Gerhard}}]{bender:94}
{Bender}, R., {Saglia}, R.~P., \& {Gerhard}, O.~E. 1994, MNRAS, 269, 785

\bibitem[{{Brainerd}(2004)}]{brainerd:04}
{Brainerd}, T.~G. 2004, in AIP Conf. Proc. 743: The New Cosmology: Conference
  on Strings and Cosmology, ed. R.~E. {Allen}, D.~V. {Nanopoulos}, \& C.~N.
  {Pope}, 129--156

\bibitem[{{Brainerd} {et~al.}(1996){Brainerd}, {Blandford}, \&
  {Smail}}]{brainerd:96}
{Brainerd}, T.~G., {Blandford}, R.~D., \& {Smail}, I. 1996, ApJ, 466, 623

\bibitem[{{Broadhurst} {et~al.}(2005){Broadhurst}, {Ben{\'{\i}}tez}, {Coe},
  {Sharon}, {Zekser}, {White}, {Ford}, {Bouwens}, {Blakeslee}, {Clampin},
  {Cross}, {Franx}, {Frye}, {Hartig}, {Illingworth}, {Infante}, {Menanteau},
  {Meurer}, {Postman}, {Ardila}, {Bartko}, {Brown}, {Burrows}, {Cheng},
  {Feldman}, {Golimowski}, {Goto}, {Gronwall}, {Herranz}, {Holden}, {Homeier},
  {Krist}, {Lesser}, {Martel}, {Miley}, {Rosati}, {Sirianni}, {Sparks},
  {Steindling}, {Tran}, {Tsvetanov}, \& {Zheng}}]{broadhurst:05}
{Broadhurst}, T., {Ben{\'{\i}}tez}, N., {Coe}, D., {Sharon}, K., {Zekser}, K.,
  {White}, R., {Ford}, H., {Bouwens}, R., {Blakeslee}, J., {Clampin}, M.,
  {Cross}, N., {Franx}, M., {Frye}, B., {Hartig}, G., {Illingworth}, G.,
  {Infante}, L., {Menanteau}, F., {Meurer}, G., {Postman}, M., {Ardila}, D.~R.,
  {Bartko}, F., {Brown}, R.~A., {Burrows}, C.~J., {Cheng}, E.~S., {Feldman},
  P.~D., {Golimowski}, D.~A., {Goto}, T., {Gronwall}, C., {Herranz}, D.,
  {Holden}, B., {Homeier}, N., {Krist}, J.~E., {Lesser}, M.~P., {Martel},
  A.~R., {Miley}, G.~K., {Rosati}, P., {Sirianni}, M., {Sparks}, W.~B.,
  {Steindling}, S., {Tran}, H.~D., {Tsvetanov}, Z.~I., \& {Zheng}, W. 2005,
  ApJ, 621, 53

\bibitem[{{dell'Antonio} \& {Tyson}(1996)}]{dell'antonio:96}
{dell'Antonio}, I.~P., \& {Tyson}, J.~A. 1996, ApJ, 473, L17+

\bibitem[{{Diego} {et~al.}(2005){Diego}, {Sandvik}, {Protopapas}, {Tegmark},
  {Ben{\'{\i}}tez}, \& {Broadhurst}}]{diego:05b}
{Diego}, J.~M., {Sandvik}, H.~B., {Protopapas}, P., {Tegmark}, M.,
  {Ben{\'{\i}}tez}, N., \& {Broadhurst}, T. 2005, MNRAS, 362, 1247

\bibitem[{{Djorgovski} \& {Davis}(1987)}]{djorgovski:87}
{Djorgovski}, S., \& {Davis}, M. 1987, ApJ, 313, 59

\bibitem[{{Dressler} {et~al.}(1987){Dressler}, {Lynden-Bell}, {Burstein},
  {Davies}, {Faber}, {Terlevich}, \& {Wegner}}]{dressler:87}
{Dressler}, A., {Lynden-Bell}, D., {Burstein}, D., {Davies}, R.~L., {Faber},
  S.~M., {Terlevich}, R., \& {Wegner}, G. 1987, ApJ, 313, 42

\bibitem[{{Duc} {et~al.}(2002){Duc}, {Poggianti}, {Fadda}, {Elbaz}, {Flores},
  {Chanial}, {Franceschini}, {Moorwood}, \& {Cesarsky}}]{duc:02}
{Duc}, P.-A., {Poggianti}, B.~M., {Fadda}, D., {Elbaz}, D., {Flores}, H.,
  {Chanial}, P., {Franceschini}, A., {Moorwood}, A., \& {Cesarsky}, C. 2002,
  A\&A, 382, 60

\bibitem[{{Faber} \& {Jackson}(1976)}]{faber:76}
{Faber}, S.~M., \& {Jackson}, R.~E. 1976, ApJ, 204, 668

\bibitem[{{Fischer} {et~al.}(2000){Fischer}, {McKay}, {Sheldon}, {Connolly},
  {Stebbins}, {Frieman}, {Jain}, {Joffre}, {Johnston}, {Bernstein}, {Annis},
  {Bahcall}, {Brinkmann}, {Carr}, {Csabai}, {Gunn}, {Hennessy}, {Hindsley},
  {Hull}, {Ivezi{\'c}}, {Knapp}, {Limmongkol}, {Lupton}, {Munn}, {Nash},
  {Newberg}, {Owen}, {Pier}, {Rockosi}, {Schneider}, {Smith}, {Stoughton},
  {Szalay}, {Szokoly}, {Thakar}, {Vogeley}, {Waddell}, {Weinberg}, {York}, \&
  {The SDSS Collaboration}}]{fischer:00}
{Fischer}, P., {McKay}, T.~A., {Sheldon}, E., {Connolly}, A., {Stebbins}, A.,
  {Frieman}, J.~A., {Jain}, B., {Joffre}, M., {Johnston}, D., {Bernstein}, G.,
  {Annis}, J., {Bahcall}, N.~A., {Brinkmann}, J., {Carr}, M.~A., {Csabai}, I.,
  {Gunn}, J.~E., {Hennessy}, G.~S., {Hindsley}, R.~B., {Hull}, C.,
  {Ivezi{\'c}}, {\v Z}., {Knapp}, G.~R., {Limmongkol}, S., {Lupton}, R.~H.,
  {Munn}, J.~A., {Nash}, T., {Newberg}, H.~J., {Owen}, R., {Pier}, J.~R.,
  {Rockosi}, C.~M., {Schneider}, D.~P., {Smith}, J.~A., {Stoughton}, C.,
  {Szalay}, A.~S., {Szokoly}, G.~P., {Thakar}, A.~R., {Vogeley}, M.~S.,
  {Waddell}, P., {Weinberg}, D.~H., {York}, D.~G., \& {The SDSS Collaboration}.
  2000, AJ, 120, 1198

\bibitem[{{Gavazzi} {et~al.}(2004){Gavazzi}, {Mellier}, {Fort}, {Cuillandre},
  \& {Dantel-Fort}}]{gavazzi:04}
{Gavazzi}, R., {Mellier}, Y., {Fort}, B., {Cuillandre}, J.-C., \&
  {Dantel-Fort}, M. 2004, A\&A, 422, 407

\bibitem[{{Geiger} \& {Schneider}(1998)}]{geiger:98}
{Geiger}, B., \& {Schneider}, P. 1998, MNRAS, 295, 497

\bibitem[{{Geiger} \& {Schneider}(1999)}]{geiger:99}
---. 1999, MNRAS, 302, 118

\bibitem[{{Ghigna} {et~al.}(1998){Ghigna}, {Moore}, {Governato}, {Lake},
  {Quinn}, \& {Stadel}}]{ghigna:98}
{Ghigna}, S., {Moore}, B., {Governato}, F., {Lake}, G., {Quinn}, T., \&
  {Stadel}, J. 1998, MNRAS, 300, 146

\bibitem[{{Ghigna} {et~al.}(2000){Ghigna}, {Moore}, {Governato}, {Lake},
  {Quinn}, \& {Stadel}}]{ghigna:00}
---. 2000, ApJ, 544, 616

\bibitem[{{Halkola} {et~al.}(2006){Halkola}, {Seitz}, \& M.}]{halkola:06}
{Halkola}, A., {Seitz}, S., \& M., P. 2006, Astro-ph, {0605470, accepted for
  publication in the MNRAS}

\bibitem[{{Hartwick} \& {Sargent}(1978)}]{hartwick:78}
{Hartwick}, F.~D.~A., \& {Sargent}, W.~L.~W. 1978, ApJ, 221, 512

\bibitem[{{Hoekstra}(2003)}]{hoekstra:03}
{Hoekstra}, H. 2003, MNRAS, 339, 1155

\bibitem[{{Hoekstra} {et~al.}(2004){Hoekstra}, {Yee}, \&
  {Gladders}}]{hoekstra:04}
{Hoekstra}, H., {Yee}, H.~K.~C., \& {Gladders}, M.~D. 2004, ApJ, 606, 67

\bibitem[{{Hudson} {et~al.}(1998){Hudson}, {Gwyn}, {Dahle}, \&
  {Kaiser}}]{hudson:98}
{Hudson}, M.~J., {Gwyn}, S.~D.~J., {Dahle}, H., \& {Kaiser}, N. 1998, ApJ, 503,
  531

\bibitem[{{Limousin} {et~al.}(2006){Limousin}, {Kneib}, {Bardeau}, {Natarajan},
  {Czoske}, {Smail}, {Ebeling}, \& {Smith}}]{limousin:06}
{Limousin}, M., {Kneib}, J.~P., {Bardeau}, S., {Natarajan}, P., {Czoske}, O.,
  {Smail}, I., {Ebeling}, H., \& {Smith}, G.~P. 2006, ArXiv Astrophysics
  e-prints

\bibitem[{{Limousin} {et~al.}(2005){Limousin}, {Kneib}, \&
  {Natarajan}}]{limousin:05}
{Limousin}, M., {Kneib}, J.-P., \& {Natarajan}, P. 2005, MNRAS, 356, 309

\bibitem[{{McKay} {et~al.}(2001){McKay}, {Sheldon}, {Racusin}, {Fischer},
  {Seljak}, {Stebbins}, {Johnston}, {Frieman}, {Bahcall}, {Brinkmann},
  {Csabai}, {Fukugita}, {Hennessy}, {Ivezic}, {Lamb}, {Loveday}, {Lupton},
  {Munn}, {Nichol}, {Pier}, \& {York}}]{mckay:01}
{McKay}, T.~A., {Sheldon}, E.~S., {Racusin}, J., {Fischer}, P., {Seljak}, U.,
  {Stebbins}, A., {Johnston}, D., {Frieman}, J.~A., {Bahcall}, N., {Brinkmann},
  J., {Csabai}, I., {Fukugita}, M., {Hennessy}, G.~S., {Ivezic}, Z., {Lamb},
  D.~Q., {Loveday}, J., {Lupton}, R.~H., {Munn}, J.~A., {Nichol}, R.~C.,
  {Pier}, J.~R., \& {York}, D.~G. 2001, Astro-ph

\bibitem[{{Merritt}(1983)}]{merritt:83}
{Merritt}, D. 1983, ApJ, 264, 24

\bibitem[{{Moore} {et~al.}(1996){Moore}, {Katz}, {Lake}, {Dressler}, \&
  {Oemler}}]{moore:96}
{Moore}, B., {Katz}, N., {Lake}, G., {Dressler}, A., \& {Oemler}, Jr., A. 1996,
  Nat, 379, 613

\bibitem[{{Moore} {et~al.}(1998){Moore}, {Lake}, \& {Katz}}]{moore:98}
{Moore}, B., {Lake}, G., \& {Katz}, N. 1998, ApJ, 495, 139

\bibitem[{{Natarajan} \& {Kneib}(1997)}]{natarajan:97}
{Natarajan}, P., \& {Kneib}, J.-P. 1997, MNRAS, 287, 833

\bibitem[{{Natarajan} {et~al.}(2002){Natarajan}, {Kneib}, \&
  {Smail}}]{natarajan:02}
{Natarajan}, P., {Kneib}, J.-P., \& {Smail}, I. 2002, ApJ, 580, L11

\bibitem[{{Natarajan} {et~al.}(1998){Natarajan}, {Kneib}, {Smail}, \&
  {Ellis}}]{natarajan:98}
{Natarajan}, P., {Kneib}, J.-P., {Smail}, I., \& {Ellis}, R.~S. 1998, ApJ, 499,
  600

\bibitem[{{Prada} {et~al.}(2003){Prada}, {Vitvitska}, {Klypin}, {Holtzman},
  {Schlegel}, {Grebel}, {Rix}, {Brinkmann}, {McKay}, \& {Csabai}}]{prada:03}
{Prada}, F., {Vitvitska}, M., {Klypin}, A., {Holtzman}, J.~A., {Schlegel},
  D.~J., {Grebel}, E.~K., {Rix}, H.-W., {Brinkmann}, J., {McKay}, T.~A., \&
  {Csabai}, I. 2003, ApJ, 598, 260

\bibitem[{{Smith} {et~al.}(2001){Smith}, {Bernstein}, {Fischer}, \&
  {Jarvis}}]{smith:01}
{Smith}, D.~R., {Bernstein}, G.~M., {Fischer}, P., \& {Jarvis}, M. 2001, ApJ,
  551, 643

\bibitem[{{Sofue} \& {Rubin}(2001)}]{sofue:01}
{Sofue}, Y., \& {Rubin}, V. 2001, ARA\&A, 39, 137

\bibitem[{{Teague} {et~al.}(1990){Teague}, {Carter}, \& {Gray}}]{teague:90}
{Teague}, P.~F., {Carter}, D., \& {Gray}, P.~M. 1990, ApJS, 72, 715

\bibitem[{{Tormen} {et~al.}(1998){Tormen}, {Diaferio}, \& {Syer}}]{tormen:98}
{Tormen}, G., {Diaferio}, A., \& {Syer}, D. 1998, MNRAS, 299, 728

\bibitem[{{Tyson} {et~al.}(1984){Tyson}, {Valdes}, {Jarvis}, \&
  {Mills}}]{tyson:84}
{Tyson}, J.~A., {Valdes}, F., {Jarvis}, J.~F., \& {Mills}, A.~P. 1984, ApJ,
  281, L59

\bibitem[{{Wilson} {et~al.}(2001){Wilson}, {Kaiser}, {Luppino}, \&
  {Cowie}}]{wilson:01}
{Wilson}, G., {Kaiser}, N., {Luppino}, G.~A., \& {Cowie}, L.~L. 2001, ApJ, 555,
  572

\bibitem[{{Zaritsky} {et~al.}(1989){Zaritsky}, {Olszewski}, {Schommer},
  {Peterson}, \& {Aaronson}}]{zaritsky:89}
{Zaritsky}, D., {Olszewski}, E.~W., {Schommer}, R.~A., {Peterson}, R.~C., \&
  {Aaronson}, M. 1989, ApJ, 345, 759

\bibitem[{{Zekser} {et~al.}(2006){Zekser}, {White}, {Broadhurst},
  {Ben{\'{\i}}tez}, {Ford}, {Illingworth}, {Blakeslee}, {Postman}, {Jee}, \&
  {Coe}}]{zekser:06}
{Zekser}, K.~C., {White}, R.~L., {Broadhurst}, T.~J., {Ben{\'{\i}}tez}, N.,
  {Ford}, H.~C., {Illingworth}, G.~D., {Blakeslee}, J.~P., {Postman}, M.,
  {Jee}, M.~J., \& {Coe}, D.~A. 2006, ApJ, 640, 639

\end{thebibliography}
\end{document}